# Measuring Zak phase in room-temperature atoms

Ruosong Mao[1,5], Xingqi Xu[1,5], Jiefei Wang[1,5], Chenran Xu[1], Gewei Qian[1], Han Cai[2,1,*], Shi-Yao Zhu[1,3] & Da-Wei Wang[1,3,4,†]

[1] *Interdisciplinary Center for Quantum Information, State Key Laboratory of Modern Optical Instrumentation, and Zhejiang Province Key Laboratory of Quantum Technology and Device, Department of Physics, Zhejiang University, Hangzhou 310027, China*

[2] *College of Optical Science and Engineering, Zhejiang University, Hangzhou, 310027, China*

[3] *Hefei National Laboratory, Hefei 230088, China*

[4] *CAS Center of Excellence in Topological Quantum Computation, Beijing 100190, China*

[5] *These authors contributed equally: Ruosong Mao, Xingqi Xu, and Jiefei Wang.*

[*] *e-mail: hancai@zju.edu.cn.*

[†] *e-mail: dwwang@zju.edu.cn.*

**Cold atoms provide a flexible platform for synthesizing and characterizing topological matter, where geometric phases play a central role. However, cold atoms are intrinsically prone to thermal noise, which can overwhelm the topological response and hamper promised applications. On the other hand, geometric phases also determine the energy spectra of particles subjected to a static force, based on the polarization relation between Wannier-Stark ladders and geometric Zak phases. By exploiting this relation, we develop a method to extract geometric phases from energy spectra of room-temperature superradiance lattices, which are momentum-space lattices of timed Dicke states. In such momentum-space lattices the thermal motion**

**of atoms, instead of being a source of noise, provides effective forces which lead to spectroscopic signatures of the Zak phases. We measure Zak phases directly from the anti-crossings between Wannier-Stark ladders in the Doppler-broadened absorption spectra of superradiance lattices. Our approach paves the way of measuring topological invariants and developing their applications in room-temperature atoms.**

Introduction

Topological matter has promising applications in noise resilient devices and quantum information processing (1,2), thanks to the robust topological response guaranteed by global geometric quantities of the Bloch bands, namely, the topological invariants (3–6). These invariants change stepwisely only when the bulk goes through a topological phase transition, which involves band gap closing and reopening. Characterizing topological invariants is a central task in synthesizing and simulating topological phases of matter. They are usually measured by the response from gapless edge states based on the bulk-edge correspondence. However, edges are not always available in atomic quantum simulators (7). On the other hand, the topological invariants are proportional to the geometric phases accumulated across a whole Brillouin zone. We can also measure the geometric phases from the bulk energy bands to obtain the topological invariants. Along this line techniques based on reciprocal-space interference (8), quench dynamics (9–12), and Hall transport (13, 14) have been developed in atomic simulators.

Previous experiments of determining geometric phases from bulk response in ultracold atoms rely on dynamic evolution or adiabatic manipulation (9–16). It has been shown that geometric phases can be obtained from the energy spectra of electrons in a constant force (15–18), which turns the Bloch energy bands into Wannier-Stark ladders (WSLs) with equidistant discrete energies. In one-dimensional (1D) systems, the displacement of the energies of the WSLs is proportional to the applied force and the positions of the Wannier centers (WCs) (19), which reflect the values of one dimensional geometric phases, i.e., the Zak phases (3) of the energy bands (20–22), as schematically illustrated in Fig. 1(a) and (b). Here we develop and implement such a spectroscopic method to retrieve Zak phases. We show that this method enables the determination of Zak phases in room-temperature atoms, which greatly improves the accessibility of topological matters and facilitates their applications.

We reconstruct the Zak phases of the Rice-Mele (RM) model through the anticrossing between the WSLs in a tight-binding lattice (24–26) of timed Dicke states (27),

$$|b(k)\rangle = \frac{1}{\sqrt{N}} \sum_m e^{ik \cdot r_m} |g_1, g_2, \ldots, b_m, \ldots, g_N\rangle, \qquad (1)$$

i.e., single-photon collective excitations of an ensemble of $N$ atoms (here $r_m$ is the position of the $m$th atom with the ground state $|g_m\rangle$ and an excited state $|b_m\rangle$), which can be created by coherent laser fields that transfer momentum $\hbar k$ to the atoms. We introduce multiple laser fields to couple $|b\rangle$ to another atomic state $|a\rangle$ such that timed Dicke states with discrete k values form a momentum-space tight-binding lattice, which is coined the superradiance lattice. When the momentum of a timed Dicke state matches that of light, directional superradiant emission of radiation can be observed (28), which provides a convenient way to measure the lattice transport. A substantial difference between the momentum-space superradiance lattice and conventional real-space lattices in solids is that the Brillouin zone (BZ) of the superradiance lattice is in real space, where atoms at different positions can be independently diagonalized by only considering the local field strengths of a spatially periodic coupling field (here a standing wave). The position of atoms plays the same role of the lattice momentum of electrons in solids. A remarkable consequence is that atoms in motion travel through the real-space BZs periodically, following the same dynamics of electrons subjected to a constant electric field. Therefore, atomic motion provides an effective electric field (or more precisely a constant force) for the excitations in superradiance lattices.

The Zak phases of a tight-binding lattice can be obtained from the WSLs when an effective force is introduced in the lattice. Since the motion of atoms provides such an effective force in superradiance lattices, we can take advantage of the thermal motion of the atoms to read out the Zak phases from the energy spectra. In particular, different velocity groups of thermal atoms provide a set of continuous values of the effective force, which results in a set of WSLs with displaced energies proportional the Zak phases (15, 16) (see Fig. 1(c)). A key to extract the Zak phases from the absorption spectra of superradiance lattices is that WSLs from the two energy bands have anti-crossings when they approach the same energy, which results in absorption peaks and dips. The Zak phases can be obtained from simple geometric relations between the locations of the anti-crossing points and the band centers. We investigate in detail two celebrated versions of the RM model, the Semenoff insulator and the Su-Shrieffer-Heeger (SSH) model (29,30). We also demonstrate the Zak phase reconstruction for general RM models. Our method of measuring Zak phase in 1D systems can be generalized to identify topological invariants in higher dimensions (31–35).

**Results**

**Experimental setup and model**

We perform the experiment with the hyperfine levels of the $^{87}$Rb D1 line in a standing-wave-coupled EIT configuration, as shown in Fig. 1(d) (see the complete setup in Supplementary Note 1). A weak probe field propagating in $x$ direction couples the ground state $|g\rangle \equiv 5^2S_{1/2}, F = 1\rangle$ to the excited state $|b\rangle \equiv |5^2P_{1/2}, F = 2\rangle$. The excited state is also coupled to a metastable state $|a\rangle \equiv |5^2S_{1/2}, F = 2\rangle$ by two strong counter-propagating light fields, forming a 1D bipartite superradiance lattice (24-26). The absorption spectra of the probe field are used to obtain the Zak phases of the superradiance lattices.

The total Hamiltonian of the superradiance lattice is $H = H_s + H_p + H_f$, where $H_s$ and $H_p$ are interaction Hamiltonians involving the coupling and probe fields, and $H_f$ is the linear potential induced by atomic motion (see Fig. 1(e)). $H_s$ is the tight-binding Hamiltonian of the RM superradiance lattices (24–26) (we set $\hbar = 1$ and see Materials and Methods),

$$H_s = \sum_j \left( \Delta_c\, \hat{a}_{2j}^\dagger \hat{a}_{2j} + \left[ \hat{a}_{2j}^\dagger \left( \Omega_1 \hat{b}_{2j+1} + \Omega_2 \hat{b}_{2j-1} \right) + \text{H.c.} \right] \right), \qquad (2)$$

where $\Omega_1$ and $\Omega_2$ are the Rabi frequencies of the co-propagating and counter-propagating coupling fields, and $\Delta_c = \nu_c - \omega_{ba}$ with $\nu_c$ being the coupling field frequency and $\omega_{ba}$ being the transition frequency between states $|a\rangle$ and $|b\rangle$. Here $\hat{d}_j^\dagger = \sqrt{1/N} \sum_m |d_m\rangle\langle g_m| \exp(ik_j x_m)$ $(d = a, b)$ is the creation operator of the timed-Dicke state (27) $|\tilde{d}_j\rangle \equiv \hat{d}_j^\dagger |g_1, g_2, \dots, g_N\rangle$ with wave vector $k_j = k_p + (j-1)k_c \approx jk_c$ ($j$ is an integer) and $k_p$ ($k_c$) being the probe (coupling) field wave vector amplitude, $m$ labels the $m$th atom at the position $x_m$, $N$ is the total number of atoms within the velocity range $[v - \Gamma/2k_c, v + \Gamma/2k_c]$ where $v$ satisfies the Maxwell distribution and $\Gamma$ is the decay rate of the state $|b\rangle$ (we neglect the decay of the hyperfine ground state $|a\rangle$). The timed Dicke state $|\tilde{b}_1\rangle$ in the superradiance lattice can be created from the ground state by $H_p = \sqrt{N}\Omega_p e^{-i\Delta_p t} \hat{b}_1^\dagger + \text{H.c.}$, where the probe detuning $\Delta_p = \nu_p - \omega_{bg}$ with $\nu_p$ being the probe field frequency and $\omega_{bg}$ being the transition frequency between states $|b\rangle$ and $|g\rangle$.

In order to clarify the effect of atomic motion, we show the contribution from atoms with different velocities in x direction. For atoms in each velocity group, the opposite

Doppler shifts of the two coupling fields lead to a linear potential (see Fig. 1(e)) in momentum-space (24, 36),

$$H_f = \delta \sum_j [2j \hat{a}_{2j}^\dagger \hat{a}_{2j} + (2j+1)\hat{b}_{2j+1}^\dagger \hat{b}_{2j+1}], \quad (3)$$

where the Doppler shift $\delta \approx k_c v \approx k_p v$ with $v$ being the velocity of the atoms in $x$ direction.

**Wannier-Stark ladders**

The energy spectrum of the Hamiltonian $H_s + H_f$ is closely related to the WCs, which are the expected positions of the Wannier functions (22) in unit cells. The WCs in the $n$th unit cell $r_\pm^{[n]}$ for the upper (+) and lower (−) energy bands of $H_s$ are related to the geometric Zak phases by (in unit of distance between neighbouring lattice sites, see Fig. 1(b)) (20, 21),

$$r_\pm^{[n]} = 2n + \theta_\pm/\pi, \quad (4)$$

where the Zak phases $\theta_\pm \equiv i \int_0^{\pi/k_c} dx \langle u_\pm(x) | \partial_x | u_\pm(x) \rangle$ with $|u_\pm(x)\rangle$ being the periodic Bloch functions of $H_s$ in real space and the integration is over the whole Brillouin zone. Therefore, the Zak phases are the fractional parts of the corresponding WCs (20, 21).

The extended Bloch energy spectra split into discrete WSLs (15, 16, 36) with energy spacing proportional to the static force $\delta$ when $H_f$ is perturbative (see Fig. 1(c)),

$$E_\pm^{[n]}(\delta) = \epsilon_\pm + r_\pm^{[n]} \delta, \quad (5)$$

where $\epsilon_\pm$ denote the energies of the Bloch band centers (bc), defined as the average band energies of $H_s$ (See Supplementary Note 2 for the derivation of Eq. (5) and discussion on its validity). From Eqs. (4) and (5), the Zak phases are obtained by $\theta_\pm = \left(\partial E_\pm^{[n]}/\partial\delta - 2n\right)\pi$.

The relation in Eq. (5) can be seen in the upper panels of Fig. 2(a,b) as functions of $\delta$ for two different RM lattices, namely, the Semenoff insulator with $\theta_- = \pi, \theta_+ = 0$ (Fig. 2(a)), and the topological phase of SSH model with half-integer Zak phases $\theta_\pm = 0.5\pi$ (Fig. 2(b)). Since the Zak phase is gauge-dependent (7), its value depends on the choice of the unit cell (8). In conventional lattices Zak phases are gauge dependent. However, in our experiments the Zak phase is an observable with a fixed gauge set by the zero-energy site of $H_f$, which is determined by the Doppler shifts of atoms. This is a significant difference between SLs and conventional lattices (3) (see Supplementary Note 3).

The color scales the from the projected density of states (PDOS) on the state $|\tilde{b}_1\rangle$, i.e., $\sum_l \delta_D(\Delta_p - E_l)|\langle\psi_l|\tilde{b}_1\rangle|^2$, where $\delta_D(\Delta_p - E_l)$ is the Dirac delta function, $|\psi_l\rangle$ and $E_l$ are the eigenstates and eigenenergies satisfying $(H_s + H_f)|\psi_l\rangle = E_l|\psi_l\rangle$ (see Materials and Methods). In the weak force regime where the coupling between the WSLs from different bands are negligible, the spectra follow the linear dependence in Eq. (5) as indicated by the dashed lines. The brightest ladders in the absorption spectra are the ones corresponding to $r_\pm^{[0]} = \theta_\pm/\pi$ in the 0th unit cell, which contains the state $|\tilde{b}_1\rangle$.

**Anti-crossing of Wannier-Stark ladders**

When a pair of WSLs from different bands ($E_-^{[n]}$ and $E_+^{[m]}$) have the same energy for a $\delta$, the interband coupling removes their degeneracy and results in an anti-crossing (37) denoted by $ac_{n,m}$. Their positions in the energy-force diagram can be estimated by the degeneracy points of the uncoupled WSLs (38) satisfying $E_-^{[n]}(\delta) = E_+^{[m]}(\delta) = \Delta_{n,m}$, where $\Delta_{n,m}$ is the probe detuning of the corresponding anti-crossing point. The values of $\Delta_{n,m}$ obtained from the experimental absorption spectra are the key to extract the Zak phases.

The optical responses (reflection and absorption) of the superradiance lattice are contributed by all atoms in Maxwell velocity distribution (39). We obtain the averaged PDOS in the lower panels of Fig. 2(a, b) from the corresponding WSLs in the upper panels by integrating $\delta$, which has a Doppler width about 500 MHz and covers all relevant values for the Zak phase reconstruction. We need to emphasize here that our scheme only requires that the velocity distribution shall be large enough to cover all the relevant anti-crossings. The Maxwell distribution of room-temperature atoms satisfies such a requirement (see the experimental spectra at different temperatures in Supplementary Note 4). The method is equally valid for other velocity distributions, as well as for cold atoms whose velocities can be precisely controlled.

Since the absorption coefficient is proportional to the PDOS (24), the anti-crossings and band centers modify the PDOS drastically and their signatures can be easily picked out in the absorption spectra. The values of $\Delta_{n,m}$ and $\epsilon_{\pm}$ are experimentally measured with the corresponding extrema in the absorption spectra (see exemplary datasets in Supplementary Note 5). As shown in Fig. 2(c) (and more examples in Fig. 4), the spectra of the two-band SLs are generally featured with four dips, of which two are associated with the band centers and the rest two are due to anti-crossings. The band centers are generally characterized by dips in the far left and far right of the spectra, owing to the Stark localization. Only in a special case with zero Zak phase, a band center is featured by a peak in Fig. 2(a) (see Supplementary Note 6). Between the two band centers, the anti-crossings of WSLs lead to dips in the spectra, reflecting the energy gaps of the anti-crossings. Since we measure the PDOS of the state $|\tilde{b}_1\rangle$ in the 0th unit cell, the major anti-crossings are associated to the Wannier functions localized in the 0th and the neighbouring −1st unit cells, i.e., $\Delta_{-1,0}$ and $\Delta_{0,-1}$.

We also show the reflection spectra in Fig. 2(c, d), which is the directional emission from the state $|\tilde{b}_{-1}\rangle$ along $-x$ direction. The reflection spectra also have features

characterizing the anti-crossing of the WSLs (e.g., the peaks of reflection spectra correspond to band centers and anti-crossings). On the other hand, they can also be used to study the lattice transport between sites $|\tilde{b}_1\rangle \to |\tilde{b}_{-1}\rangle$ for lattices in different topological phases, which is out of the scope of the current paper.

**Zak phase measurement**

In order to measure the Zak phases, we shall quantify the common features in the absorption spectra of lattices with the same Zak phase. In Fig. 3(a), we maintain $\Omega_1 = \Omega_2 = 120$ MHz and decrease $\Delta_c$ from bottom to top. The Zak phase is the same but the coupling between WSLs increases to widen the anti-crossing gap. We locate the anti-crossing points with its normalized energy,

$$R_{n,m} = (\Delta_{n,m} - \epsilon_-)/(\epsilon_+ - \epsilon_-). \qquad (6)$$

On the other hand, according to the geometry of WSLs in the $\Delta_p$-$\delta$ diagram, $R_{n,m}$ is approximately the normalized ratio of WCs between the WSLs,

$$R_{n,m} \approx \frac{r_-^{[n]}}{r_-^{[n]} - r_+^{[m]}} = \frac{2n\pi + \theta_-}{(2n\pi + \theta_-) - (2m\pi + \theta_+)}. \qquad (7)$$

In Fig. 3(c), we obtain $R_{0,-2} \approx 1/5$, $R_{0,-1} \approx 1/3$. We solve two equations of $\theta_\pm$ from the values of $R_{0,-1}$ and $R_{0,-2}$ and conclude that $\theta_- \approx \pi$ and $\theta_+ \approx 0$, as shown in Fig. 3(e).

For the SSH models, we keep $\Delta_c = 0$ and tune the Rabi frequencies of the two coupling fields from almost dimerization to the topological phase transition point. The Zak phases are maintained the same while the anti-crossing energy gaps increase from top to bottom in Fig. 3(b). The measured $R_{n,m}$ in Fig. 3(d) agree well with their expected values and the reconstructed Zak phases are obtained, $\theta_\pm \approx 0.5\pi$, as shown in Fig. 3(f).

For a general RM Hamiltonian, i.e., $H_s$ with $\Omega_1 \neq \Omega_2$ and $\Delta_c \neq 0$, the Zak phases are neither integers nor half-integers. Along the yellow line in the phase diagram in Fig. 4(a), we measure the absorption spectra to obtain $R_{0,-1}$ and $R_{-1,0}$ for each coupling field detuning $\Delta_c$, as shown in Fig. 4(b), and accordingly reconstruct the Zak phases in Fig. 4(c), in comparison with the theoretical prediction as indicated by the dashed lines. As an example, we show the WSLs with $\theta_- = 0.3\pi$, $\theta_+ = 0.7\pi$ in Fig. 4(d) and with $\theta_- = 0.4\pi$, $\theta_+ = 0.6\pi$ in Fig. 4(e). The corresponding absorption and reflection spectra are plotted in Fig. 4(f) and (g), respectively.

**Discussions**

We realize the spectroscopic reconstruction of the Zak phases of momentum-space superradiance lattices. Without trapping atoms or controlling their velocities (8,9,10), we take advantage of the atomic thermal motion (40,41) to extract geometric phases from the anti-crossings of the WSLs. Therefore, our result pushes forward the room-temperature quantum simulation of topological phases. Meanwhile, it also paves a way for application of topological physics in optical devices that operate at ambient temperature.

The deviation between the reconstructed Zak phases and the theoretical prediction can be attributed to the following two reasons. First, when the band gap is small, the strong coupling between WSLs leads to a wide anti-crossing energy gap, such that the energy dip does not accurately reflect the location of the anti-crossing (see $\Delta_c \rightarrow 0$ for the Semenoff insulator in Fig. 3(e)). Second, the competition between $H_f$ and $H_s$ induces a systematic error even for small couplings between WSLs. For the SSH model in Fig. 3(f), the force required for the two major anti-crossings is $\delta \approx \pm(\epsilon_+ - \epsilon_-)/2$. The induced potential energy between neighbouring sites is comparable to the hopping strength in lattices, such that $H_f$ cannot be treated as a perturbation and the slope of WSLs $\partial E_\pm^{[n]}/\partial\delta$ deviates from $r_\pm^{[n]}$ (see Supplementary Note 3). The consequence is that the energy of the anti-crossing $\Delta_{0,-1}$ ($\Delta_{-1,0}$) is always lower (higher) than the degeneracy point predicted by the linear approximation in Eq. (5) (see the difference between spectra extrema $\Delta_{n,m}$ and crossing points of blue dotted lines in Fig. 2(a, b) and Fig. 4(d, e)), leading to a systematic error in determining the Zak phases.

To improve the accuracy in extracting the spectroscopic features of Zak phases, we are developing a spectral hole burning technique to map out the two-dimensional velocity-dependent absorption spectra of WSLs, as shown in upper panels of Fig. 2(a). By using a narrow linewidth saturation field that couples the ground state to an ancillary state, we can selectively bleach the ground state population of atoms with a certain velocity. By comparing the bleached and unbleached absorption spectra, the contribution from atoms with that velocity is obtained.

Our scheme can be generalized to measure multipole moments of higher-order topological insulators (34,35) by detecting the slopes of WSLs (18), and to measure Chern numbers by counting the Zak phase winding. In the current framework, two or higher dimensional superradiance lattices (16) can be synthesized by introducing more coupling fields (42–45), as well as in photonic lattices (46) and synthetic dimensions (47,49). We can use three coupling fields to form a 2D interference pattern in $x$-$y$ plane (Fig. 5(a)), illustrating the BZ of a momentum-space honeycomb superradiance lattices (42). We can identify the Chern number $C_\pm$ of the upper (+) and lower (-) bands from the winding number of the one-dimensional Zak phase along the perpendicular dimension (16,25-33),

$$C_\pm = \frac{1}{2\pi}\int_0^L dy \frac{\partial\theta_\pm(y)}{\partial y}, \qquad (8)$$

where $\theta_\pm(y)$ is the Zak phase along the $x$-axis cut of the 2D BZ with a fixed $y$ and $L$ is the length of real-space BZ along the $y$-axis. In Fig. 5(b), we schematically show that $\theta_\pm(y)$ can be measured from the absorption spectra of a probe field with a beam size much smaller than $L$. In order to suppress the paraxial diffraction, we need to ensure $L$ is much larger than the wavelength by minimizing the angle between the two copropagating coupling fields. After collecting the $\theta_\pm(y)$, we determine the Chern number of the 2D superradiance lattices by counting how many times it winds within $L$, as shown in Fig. 5(c).

With the ability of measuring geometric phases in SLs, a promising direction in the next stage is to introduce interactions between atoms, e.g., by using Rydberg states (50). It is particularly interesting to notice that the short-range interaction in real space has long-range effect in momentum-space SLs, which is difficult to realize in real-space lattices (51).

In conclusion, we develop a method of reconstructing Zak phases from the anti-crossings of WSLs by measuring the Doppler-broadened absorption spectra of room-temperature superradiance lattices. This method can be directly generalized to measure Chern numbers by counting the Zak phase windings in 2D lattices (16) by introducing more coupling fields (42–45). Our method can also be implemented in cold atoms (25) by controlling the atomic velocity to track the peak shifting, as sketched in Fig. 1(c). We can also use the hole burning technique to obtain the WSL of atoms with different velocities. Our results pave the way to detect multipole moments in higher-order topological insulators (18,34,35).

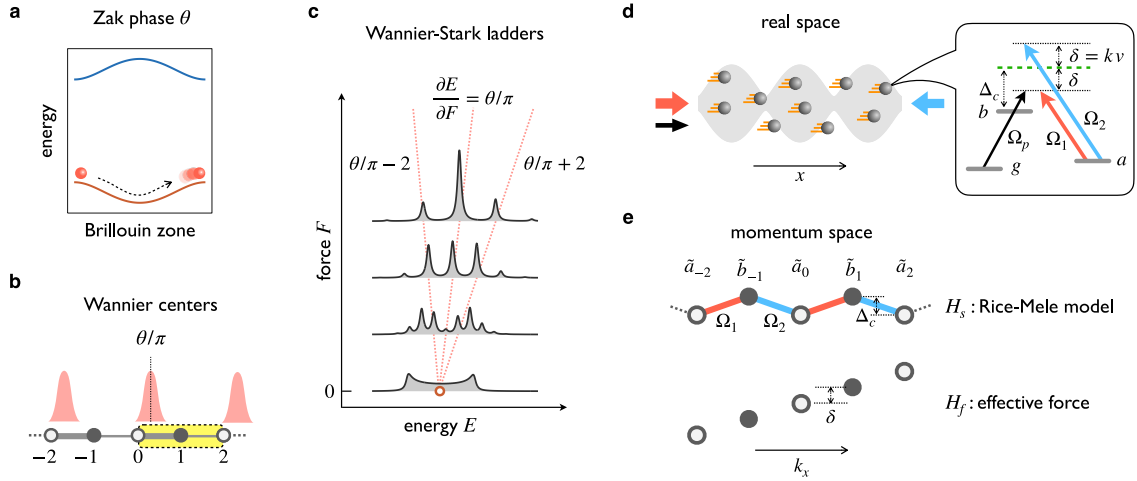

Figure 1 **Relation between Zak phases and Wannier-Stark ladders in superradiance lattices.** (a) The Zak phase as an accumulated geometric phase of a particle adiabatically driven across a whole Brillouin zone. (b) The Wannier function in each unit cell. The zeroth unit cell is highlighted with yellow color. The positions of the Wannier centers are $2n + \theta/\pi$. (c) Schematic WSLs shown by the projected density of states with different forces $F = \delta$. The three dotted lines connect the 0th and $\pm$1st WSLs. Quantum transport package Nanoskim (23) is used in the calculation. (d) Schematic configuration of the light fields. Inset: the coupling between the light fields and the atomic levels in the reference frame of the atom with the Rabi frequency $\Omega_i$ ($i = 1,2,p$), the coupling field detuning in the lab reference frame $\Delta_c$, and the Doppler shift $\delta$. The shaded area indicates the envelope of the standing wave coupling field and the balls indicate atoms with velocity $v$. (e) The momentum-space superradiance lattice with tight-binding Hamiltonian $H_s$ (upper) and a linear potential $H_f$ (lower). $\tilde{d}_j$ ($d = a, b$) are the timed Dicke states.

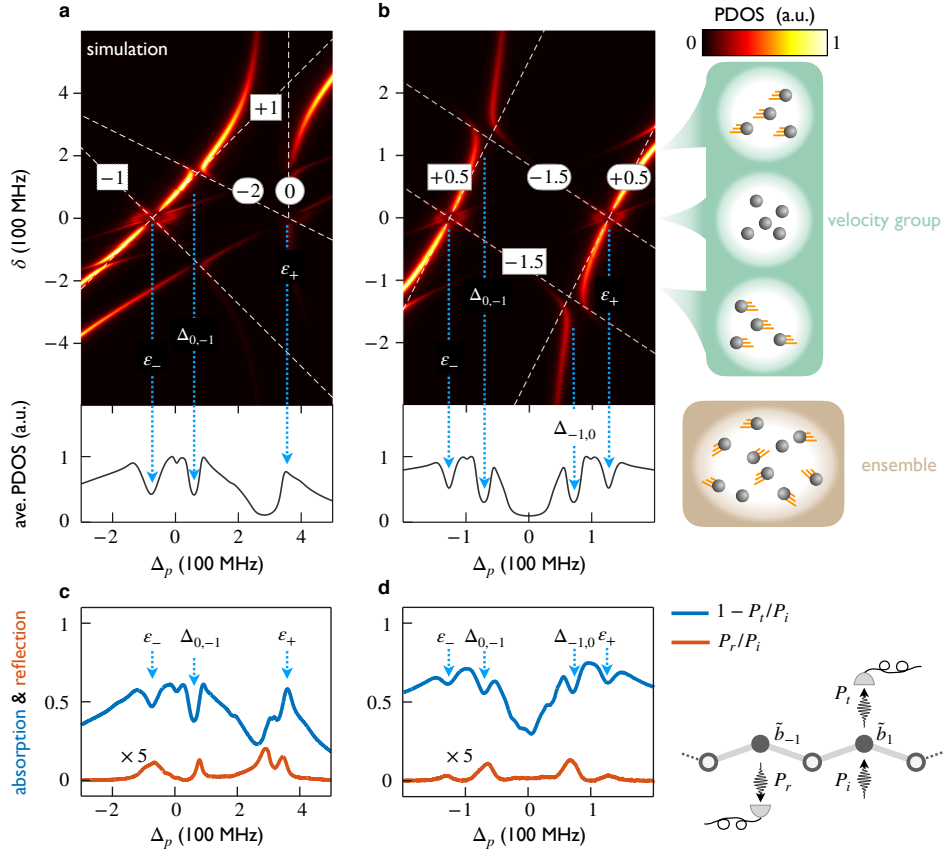

Figure 2 **Wannier-Stark ladders and the absorption spectra.** (a)-(b) The upper panels are numerical simulation of the PDOS as functions of the Doppler shift $\delta$ and probe detuning $\Delta_p$. The averaged PDOS in the lower panels are obtained from upper panels by integrating $\delta$ (in Maxwell distribution with FWHM 500 MHz). (c)-(d) The experimental data of the absorption $(1 - P_t/P_i)$ and reflection $(P_r/P_i)$ spectra, where $P_i$, $P_t$, and $P_r$ are the power of the incident, transmitted, and reflected probe fields, respectively. (a) and (c) The Semenoff insulator with $\Omega_1 = \Omega_2 = 120$ MHz and $\Delta_c = 298$ MHz. The SSH model ($\Delta_c = 0$) with (b) and (d) $\Omega_1 = 125$ MHz, $\Omega_1/\Omega_2 = 5.3$. The white dashed lines indicate the uncoupled WSLs in Eq. (4). The highlighted numbers denote the values of $r_-^{[n]}$ (square) and $r_+^{[m]}$ (round) of the corresponding WSLs. Both in the simulated averaged PDODs and the measured absorption spectra, the dips and peaks capture the band centers (denoted by $\epsilon_\pm$) and anti-crossing points (denoted by $\Delta_{n,m}$), where the blue dotted lines are used to guide eyes. The arrows point to the local extrema of the corresponding spectral features.

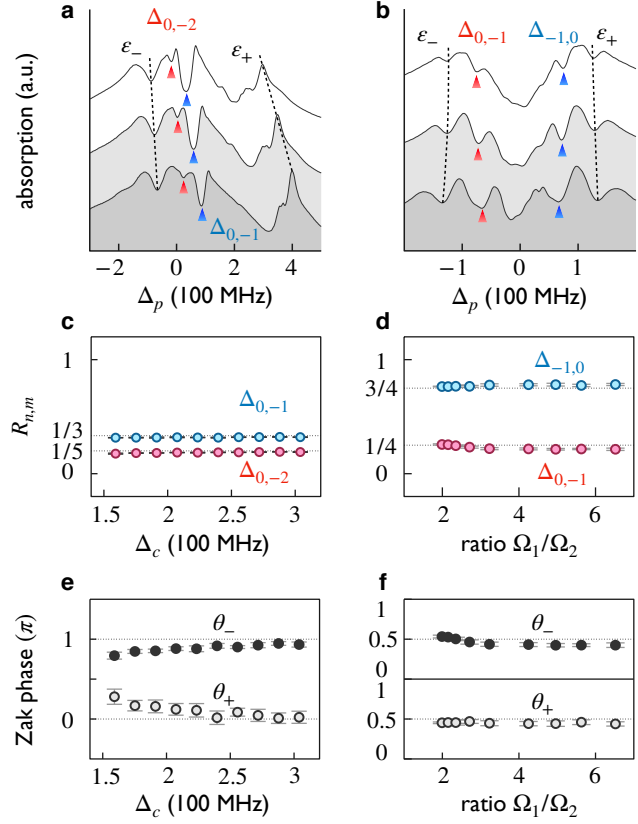

Figure 3 **Zak phase reconstruction from the absorption spectra.** The experimental data of the absorption spectra for (a) the Semenoff insulator superradiance lattices with $\Omega_{1,2} = 120$ MHz and $\Delta_c = 174$, 232, and 298 MHz from top to bottom, and (b) the SSH superradiance lattices with $\Delta_c = 0$, $\Omega_1 = 118$ MHz, and $\Omega_1/\Omega_2 = 6.52$, 4.25, and 2.15 from top to bottom. We use the marked extrema $\epsilon_\pm$ ($\Delta_{n,m}$) in absorption peaks and dips to locate the band centers (anti-crossings). The measured $R_{n,m}$ (points) compared with the normalized ratio of WCs (dashed lines) for (c) the Semenoff insulator and (d) the SSH superradiance lattices, from which we reconstruct the Zak phases $\theta_\pm$ in (e) and (f). Error bars are obtained from four independent data sets (see WSLs and more absorption spectra in Supplementary Note 7).

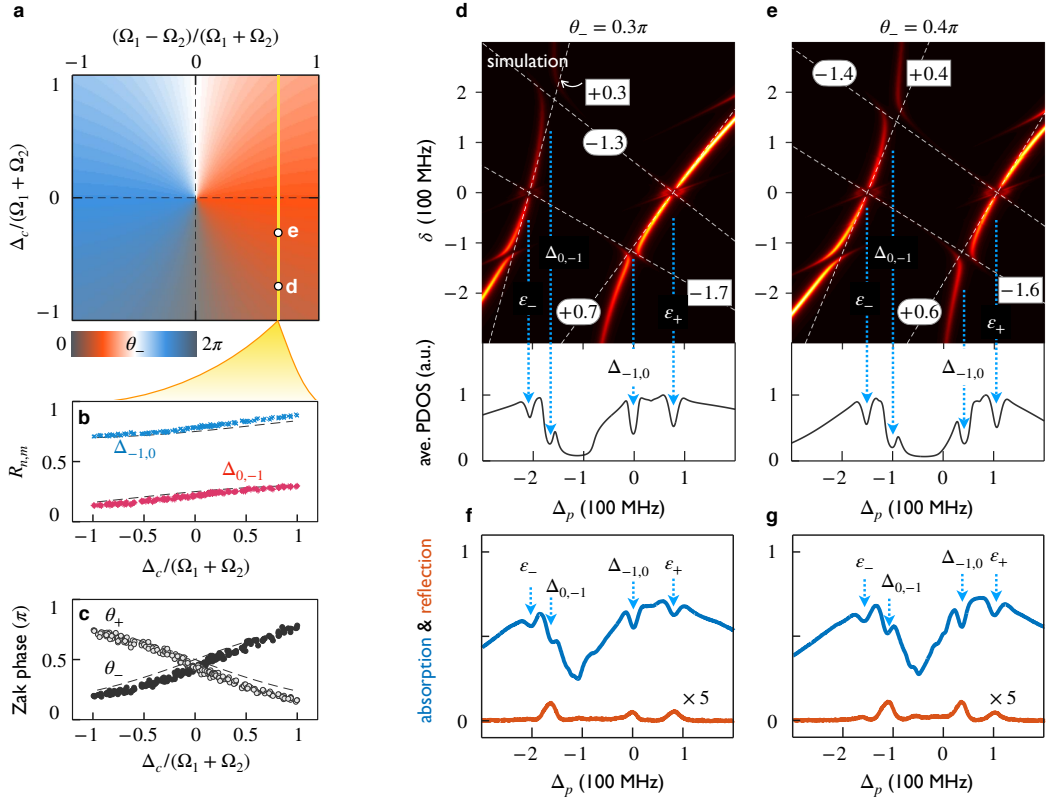

Figure 4 **Zak phase reconstruction of Rice-Mele superradiance lattices.** (a) The Zak phase diagram of the RM model as a function of $\Delta_c$ and $\Omega_1 - \Omega_2$. (b) Locations of the anti-crossing points and (c) values of the Zak phases are measured along the yellow line in (a) with $\Omega_1 = 125$ MHz and $\Omega_1/\Omega_2 = 5.3$, compared with their theoretical values (dashed lines). The plots contain 200 data sets. (d) and (e) The WSLs and the absorption/reflection spectra with $\theta_- = 0.3\pi$, $\theta_+ = 0.7\pi$, $\Delta_c = -108$ MHz. (f) and (g) The WSLs and the absorption/reflection spectra with $\theta_- = 0.4\pi$, $\theta_+ = 0.6\pi$, $\Delta_c = -45$ MHz.

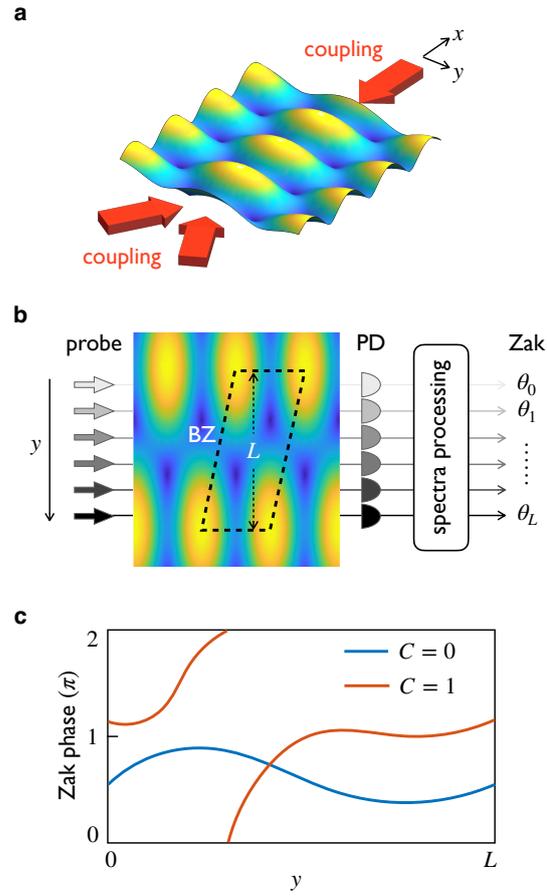

Figure 5 **Proposal of measuring Chern numbers from Zak phases.** (a) 2D interference pattern in $x$-$y$ plane induced by three coupling fields, illustrating the Brillouin zone of a superradiance lattice Haldane model. (b) Measurement of the position dependent Zak phases. For each fixed $y$ position, the $\theta_\pm(y)$ along the $x$ direction is obtained from the absorption spectrum of the corresponding probe field. (c) The winding number of $\theta_\pm(y)$ from $y = 0$ to $L$ is related to the Chern number $C$, where $L$ is the BZ width projected on $y$-axis.

**Materials and methods**

**Effective Hamiltonian**

Here we derive the effective Hamiltonian $H = H_s + H_f + H_p$. It is convenient to work with the master equation in the inertial reference frame of the moving atoms. The Rabi frequencies of the two plane components of the coupling field are $\Omega_1$ ($x$ directional) and $\Omega_2$ ($-x$ directional), and the corresponding Doppler shifted frequencies are $v_c - \delta$ and $v_c + \delta$, respectively. For the probe field propagating along $x$ direction, the Rabi frequency and Doppler shifted frequency are $\Omega_p$ and $v_p - \delta$, respectively. Here we define the Doppler shift $\delta \approx k_c v \approx k_p v$, where $k_{p(c)}$ is the wave vector amplitude of the probe (coupling) field. The original Hamiltonian of an ensemble of Λ-type three-level atoms in EIT is written as,

$$H' = \omega_{bg}|b\rangle\langle b| + \omega_{ag}|a\rangle\langle a|$$
$$+ \left(\Omega_1 e^{-i(v_c-\delta)t+ik_c x}|b\rangle\langle a| + \Omega_2 e^{-i(v_c-\delta)t-ik_c x}|b\rangle\langle a| + \text{H.c.}\right)$$
$$+ \left(\Omega_p e^{-i(v_p-\delta)t+ik_p x}|b\rangle\langle g| + \text{H.c.}\right). \quad (9)$$

In order to eliminate the dynamic phase factors, we transform the Hamiltonian into the interaction picture,

$$V = U^{-1}H'U - S = \Delta_c|a\rangle\langle a| + \left(\Omega_p e^{-i(\Delta_p-\delta)t+ik_p x}|b\rangle\langle g| + \text{H.c.}\right)$$
$$+ \left(\Omega_1 e^{i\delta t+ik_c x}|b\rangle\langle a| + \Omega_2 e^{-i\delta t-ik_c x}|b\rangle\langle a| + \text{H.c.}\right) \quad (10)$$

where $U = \exp(-iSt)$ and $S = \omega_{bg}|b\rangle\langle b| + (\omega_{bg} - v_c)|a\rangle\langle a|$.

In the experiments, the powers of the coupling fields are much larger than that of the probe field, and the probe field is far below the saturation strength, i.e., $\Omega_{1,2}, \Gamma \gg \Omega_p$.

Therefore, we only keep the first order of $\Omega_p$ but keep all orders of $\Omega_{1,2}$. The relevant dynamical equations that govern the evolution of coherence $\rho_{ag}$ and $\rho_{bg}$ are

$$\dot{\rho}_{ag} = -i[(\Omega_1 e^{-i\delta t - ik_c x} + \Omega_2 e^{i\delta t + ik_c x})\rho_{bg} + \Delta_c \rho_{ag}] - \gamma_a \rho_{ag},$$
$$\dot{\rho}_{bg} = -i[(\Omega_1 e^{i\delta t + ik_c x} + \Omega_2 e^{-i\delta t - ik_c x})\rho_{ag} + \Omega_p e^{-i(\Delta_p - \delta)t + ik_p x}] - \gamma_{bg} \rho_{bg}, \quad (11)$$

where $\gamma_{bg} = \Gamma/2 + \gamma_b$ is the decoherence rate of $\rho_{bg}$ and $\gamma_i$ is the dephasing rate of the level $|i\rangle$. The general solutions are assumed as

$$\rho_{ag} = \sum_j \rho_{ag}^{[2j]} e^{ik_{2j}x'},$$
$$\rho_{bg} = \sum_j \rho_{bg}^{[2j+1]} e^{ik_{2j+1}x'}, \quad (12)$$

where $k_j = k_p + (j-1)k_c \approx jk\_c$, $x' = x + \delta/k_c t = x + vt$ being the position of atoms in motion. In the weak excitation limit $\rho_{ag}^{[2j]}, \rho_{bg}^{[2j+1]} \ll 1$, the wavefunction of a single atom at $x_m$ is approximately $|\Psi_m\rangle \approx \rho_{ag}|a_m\rangle + \rho_{bg}|b_m\rangle + |g_m\rangle$. Therefore, the wavefunction of the whole atomic ensemble reads

$$|\Psi\rangle = \prod_m \otimes |\Psi_m\rangle$$
$$\approx \sum_m \sum_j \rho_{ag}^{[2j]} e^{ik_{2j}x'_m} |g_1 g_2 \ldots a_m \ldots g_N\rangle$$
$$+ \sum_m \sum_j \rho_{bg}^{[2j+1]} e^{ik_{2j+1}x'_m} |g_1 g_2 \ldots b_m \ldots g_N\rangle + |G\rangle$$
$$= \sum_j (\alpha_{2j} \hat{a}_{2j}^\dagger + \beta_{2j+1} \hat{b}_{2j+1}^\dagger + 1)|G\rangle, \quad (13)$$

where we use the definition of the collective ground state $|G\rangle = |g_1 g_2 \ldots g_N\rangle$ and the probability amplitudes of the timed Dicke states $\alpha_{2j} = \sqrt{N}\rho_{ag}^{[2j]}$, $\beta_{2j+1} = \sqrt{N}\rho_{bg}^{[2j+1]}$. Combining Eq. (10-12), we write the dynamic equation formally

$$i\frac{d}{dt}|\Psi\rangle = (H - i\hat{\gamma})|\Psi\rangle, \qquad (14)$$

with the effective Hamiltonian,

$$H = \sum_j \left((\Delta_c + 2j\delta)\hat{a}^\dagger_{2j}\hat{a}_{2j} + \delta(2j+1)\hat{b}^\dagger_{2j+1}\hat{b}_{2j+1}\right)$$
$$+\left[\Sigma_j\left(\Omega_1\hat{a}^\dagger_{2j}\hat{b}_{2j+1} + \Omega_2\hat{a}^\dagger_{2j}\hat{b}_{2j-1}\right) + \sqrt{N}\Omega_p e^{-i\Delta_p t}\hat{b}^\dagger_1 + \text{H.c.}\right],$$

and the dissipation operator,

$$\hat{\gamma} = \Sigma_j\left(\gamma_a\hat{a}^\dagger_{2j}\hat{a}_{2j} + \gamma_{bg}\hat{b}^\dagger_{2j+1}\hat{b}_{2j+1}\right).$$

**Absorption and PDOS**

The induced polarization of the state $|\Psi\rangle$ in Eq. (12) is defined as

$$P = \langle\Psi|er|\Psi\rangle = \sum_j \sum_m \mu \rho_{bg}^{[2j+1]} e^{ik_{2j+1}x'_m} + \text{c.c.}, \quad (15)$$

where $\mu = \langle g|er|b\rangle$ is the single atom dipole moment. Since the atoms are homogeneously distributed, the polarization density as a function of position reads

$$P(x) = n \sum_j \rho_{bg}^{[2j+1]} e^{ik_{2j+1}x'} + \text{c.c.}, \quad (16)$$

where $n$ is the atomic density. Therefore, the susceptibility of the atoms is

$$\chi = \frac{P(x)}{\epsilon_0 E_p e^{ik_p x'}} = \sum_j \chi^{[j]} e^{ik_{2j}x'}. \quad (17)$$

The optical absorption coefficient $A$ is related to the 0th-order component (linear) of the susceptibility in Eq. (17), and further connected to projected density of states (PDOS) of the superradiance lattices

$$\begin{aligned}
A &\propto \text{Im}\chi^{[0]} \propto \text{Im}\beta_1 \\
&= \text{Im}\left\langle \tilde{b}_1 \left| \frac{\sqrt{N}\Omega_p}{(\Delta_p + i\gamma_{bg}) - H_f(v) - H_s} \right| \tilde{b}_1 \right\rangle \\
&= \text{Im} \sum_l \frac{|\langle\psi_l(v)|\tilde{b}_1\rangle|^2}{(\Delta_p + i\gamma_{bg}) - E_l(v)} \\
&= -\sum_l \delta_D\big(\Delta_p - E_l(v)\big) |\langle\psi_l(v)|\tilde{b}_1\rangle|^2, \quad (18)
\end{aligned}$$

where we take advantage of the Green's function approach (52–54). Here $H_f$, $|\psi_l\rangle$, and $E_l$ are functions of $v$. The power of the transmitted probe field is $P_t = P_i e^{-As}$, where $s$ is the length of the Rb vapor cell.

**Acknowledgements**

The authors thank Luqi Yuan for useful discussions. This work was supported by the National Natural Science Foundation of China (Grants No. 11874322, No. 11934011 and No. U21A20437), the National Key Research and Development Program of China (Grants No. 2019YFA0308100, No. 2018YFA0307200 and No. 2017YFA0304202), Zhejiang Province Key Research and Development Program (Grant No. 2020C01019), the Strategic Priority Research Program of Chinese Academy of Sciences (Grant No. XDB28000000), the Fundamental Research Funds for the Central Universities and Innovation Program for Quantum Science and Technology (Grant No. 2021ZD0303200). We gratefully acknowledge HZWTECH for providing computation facilities.

**Author contribution**

H.C. and D.W.W. conceived the project. H.C. and X.X. designed the experiment. J.W., X.X., C.X., and G.Q. built the experimental setup and carried out the measurement. R.M. and H.C. did numerical simulation. R.M., J.W., X.X. and H.C. performed data analysis. D.W.W. and S.Y.Z. supervised the research. H.C. and D.W.W. wrote the manuscript with comments and contributions from all authors.

**Conflict of Interest statement**

The authors declare no competing interests.